\documentclass[sigconf]{acmart} 
\usepackage{microtype}
\usepackage{graphicx}
\usepackage{subcaption}
\usepackage{booktabs} 
\usepackage{multirow}
\usepackage{hyperref}
\usepackage{amsmath}
\usepackage{mathtools}
\usepackage{amsthm}
\usepackage[table]{xcolor}
\AtBeginDocument{%
  }

\copyrightyear{2026}
\acmYear{2026}
\setcopyright{cc}
\setcctype{by}
\acmConference[RecSys '26]{20th ACM Conference on Recommender Systems}{September 27-October 02, 2026}{Minneapolis, MN, USA}
\acmBooktitle{20th ACM Conference on Recommender Systems (RecSys '26), September 27-October 02, 2026, Minneapolis, MN, USA}
\acmDOI{10.1145/3773078.3831869}
\acmISBN{979-8-4007-2284-4/2026/09}

\begin{document}

\title[CCN: A Comprehensive Comparison Network for Route Recommendation]{Towards Full Candidate Interaction: A Comprehensive Comparison Network for Better Route Recommendation}


\author{Hanyu Guo}
\orcid{0009-0007-8877-5474}
\authornote{These authors contributed equally to this research.}
\affiliation{%
  \institution{AMAP, Alibaba Group}
  \city{Beijing}
  \country{China}
}
\email{guohanyu.ghy@alibaba-inc.com}

\author{Chao Chen}
\orcid{0009-0000-0586-6207}
\authornotemark[1]
\affiliation{%
  \institution{AMAP, Alibaba Group}
  \city{Beijing}
  \country{China}
}
\email{cc201598@alibaba-inc.com}

\author{Longfei Xu}
\orcid{0009-0005-6973-2749}
\authornotemark[1]
\affiliation{%
  \institution{AMAP, Alibaba Group}
  \city{Beijing}
  \country{China}
}
\email{longfei.xl@alibaba-inc.com}

\author{Chengzhang Wang}
\orcid{0009-0000-0229-4653}
\affiliation{%
  \institution{AMAP, Alibaba Group}
  \city{Beijing}
  \country{China}
}
\email{wangchengzhang.wcz@alibaba-inc.com}

\author{Kaikui Liu}
\orcid{0009-0000-4443-1353}
\affiliation{%
  \institution{AMAP, Alibaba Group}
  \city{Beijing}
  \country{China}
}
\email{damon@alibaba-inc.com}

\author{Xiangxiang Chu}
\orcid{0000-0003-2548-0605}
\affiliation{%
  \institution{AMAP, Alibaba Group}
  \city{Beijing}
  \country{China}
}
\email{cxxgtxy@gmail.com}

\renewcommand{\shortauthors}{Hanyu Guo, Chao Chen, Longfei Xu et al.}

\begin{abstract}
    We argue that the decision-making essence of route recommendation is comparative judgment: users choose a route because it is better than alternatives in specific aspects. The decision-critical information resides in segment-level spatial differences of non-overlapping parts between routes, which is irreversibly lost through item-level feature aggregation. Existing methods, whether attention-based or pairwise ranking approaches, follow an item-first paradigm that can only infer pairwise relations indirectly from individual route representations. To address this, we propose the Comprehensive Comparison Network (CCN), which inverts the information flow by constructing explicit comparison features from non-overlapping segments between route pairs and reasoning directly in the pairwise space. CCN introduces a Comprehensive Comparison Block that enables context-aware pairwise reasoning, where the comparison between two routes is informed by how both compare against all other candidates. We further develop an interpretable Pair Scoring Network that decomposes pairwise preferences into independent physical fields, providing field-level explanations for route selection. CCN has served as the production ranking model in Amap for over two years, achieving 85.70\% offline route-trajectory coverage rate and +1.2\% online improvement over the previous production model. We also release a large-scale route recommendation dataset comprising 175 million users, 512 million samples, and 6 billion routes across 370 cities.\footnote{Dataset: \url{https://huggingface.co/datasets/GD-ML/CCN}}
\end{abstract}

\ccsdesc[500]{Information systems~Retrieval models and ranking}
\ccsdesc[500]{Information systems~Learning to rank}

\keywords{Recommendation system, Route recommendation, Pairwise Interaction, Interpretability}
\maketitle

\section{Introduction}
In large-scale navigation platforms, route recommendation operates as a cascading pipeline: a recall stage generates diverse candidate routes via classical planning algorithms \cite{hart1968formal, abraham2013alternative}, a pre-ranking stage eliminates clearly dominated alternatives, and a final ranking stage applies expressive models with rich features to identify the optimal route. This work targets the final ranking stage.

\begin{figure}[t]
    \centering
    \includegraphics[width=0.48\textwidth]{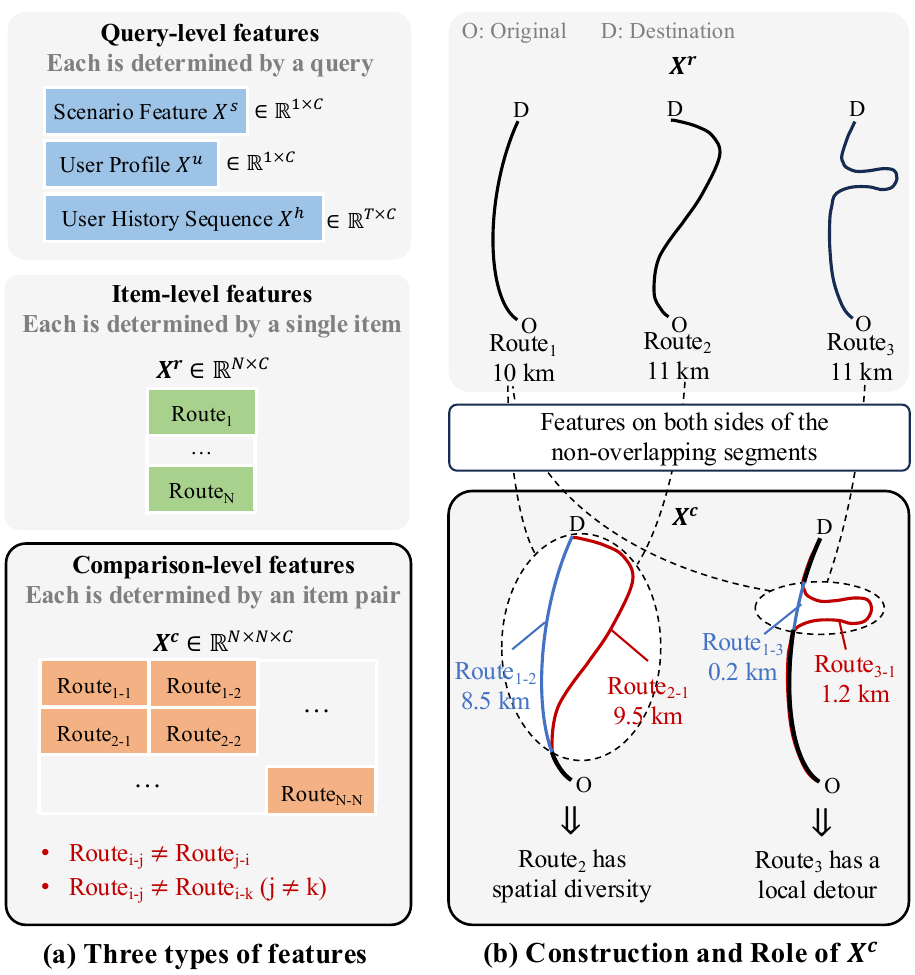}
    \caption{(a) Three types of features in recommendation systems. (b) Item-level aggregated features only reveal coarse differences, while comparison-level features from non-overlapping segments distinguish spatial diversity from local detours.}
\Description{Illustration comparing item-level, user-level, and comparison-level features, showing that non-overlapping route segments preserve spatial differences that aggregate item-level features cannot distinguish.}
    \label{fig:1}
\end{figure}

We argue that the decision-making essence of route recommendation is \textbf{comparative judgment}: users select a route not because it is independently ``good,'' but because it is \textit{better than alternatives in specific aspects}. As illustrated in Figure~\ref{fig:1}, item-level aggregated statistics such as total distance or ETA capture only coarse differences between routes. The truly decision-critical information, namely \textit{where} and \textit{why} routes diverge, resides in the segment-level spatial structure of their non-overlapping parts. Such comparison-level information is irreversibly lost through item-level aggregation.

In practice, each route is conventionally represented as a fixed-length feature vector aggregated over its constituent road segments. Existing recommendation methods, whether attention-based models \cite{kang2018self, sun2019bert4rec, zhou2018deep, zhou2019deep} or pairwise and listwise ranking methods \cite{pei2019personalized, ai2018learning, bello2018seq2slate}, naturally operate on these item-level representations and infer pairwise relations only indirectly through similarity-based attention or post-hoc score differences. This item-first paradigm presents two inherent limitations for route recommendation. First, aggregated features discard the segment-level spatial information essential for distinguishing meaningful diversity from detours, as shown in Figure~\ref{fig:1}(b). Second, even when attention mechanisms enable interactions across all candidates at the item level, the influence of other alternatives on pairwise preferences is only implicit, mediated through updated item representations rather than modeled directly in the comparison space. An intuitive alternative would be to represent each route as a sequence of road segment IDs and model inter-route differences through cross-sequence attention. However, this is impractical at scale: real-world road networks contain hundreds of millions of segments with extreme long-tail distributions, and routes vary significantly in length, making both embedding convergence and cross-route segment alignment prohibitively expensive. Even if feasible, such approaches still require the model to implicitly discover pairwise differences rather than operating on explicit comparative signals.

To address these limitations, we propose the \textbf{Comprehensive Comparison Network (CCN)}, which inverts the conventional information flow by constructing explicit comparison features as primary input and reasoning directly in the pairwise space. Rather than representing each route as a full sequence of per-segment features, CCN aggregates physical attributes at the granularity of non-overlapping segments between route pairs, preserving the spatial difference information lost by item-level aggregation while avoiding the variable-length alignment and sparsity challenges of segment-level representations. CCN introduces two core innovations. The first is \textit{Comparison-level features} derived from non-overlapping segments between route pairs, forming an $N \times N \times F$ comparison tensor as explicit, fixed-dimensional pairwise input. The second is a \textit{Comprehensive Comparison Block} that enables context-aware pairwise reasoning: the comparison between routes $i$ and $j$ is informed by how both compare against all other candidates, capturing the full competitive context of the candidate set.

Furthermore, we develop an interpretable \textit{Pair Scoring Network (PSN)} that decomposes pairwise preferences along independent physical dimensions such as ETA, toll, and road capacity, providing field-level explanations for route selection. CCN with PSN has served as the production ranking model in Amap for over two years. We also contribute a large-scale route recommendation dataset comprising 175 million users, 512 million samples, and 6 billion routes across 370 cities, with ground-truth labels derived from actual user navigation trajectories. Our contributions are as follows:
\begin{itemize}
\item We identify comparative judgment as the decision-making essence of route recommendation and propose a comparison-first paradigm that constructs explicit comparison tensors as primary input, preserving spatial differences lost by aggregation without segment-level embedding learning.
\item We design the Comprehensive Comparison Block for context-aware pairwise reasoning, where each pair's representation is informed by the full candidate set, going beyond independent pairwise scoring.
\item We develop an interpretable Pair Scoring Network that decomposes pairwise preferences into independent physical fields, enabling field-level explanations for route selection.
\item We release a large-scale route recommendation dataset and validate CCN through over two years of production deployment serving hundreds of millions of users.
\end{itemize}

\section{Related Work}
\subsection{Route Planning and Recommendation}
Route planning is a classical graph-search problem that computes optimal paths over road networks based on criteria such as shortest distance or minimum travel time \cite{hart1968formal}. Extensions to alternative route planning generate multiple diverse candidates through penalty-based, Pareto, or diversified search strategies \cite{abraham2013alternative, luo2022diversified}. While effective for individual path queries, these algorithmic approaches face scalability challenges when serving millions of concurrent requests in industrial navigation platforms.

Route recommendation addresses this gap by decoupling path generation from path selection through a recall-and-rank framework \cite{zhang2024survey, NEURIPS2021_b922ede9}: the recall stage employs efficient planning algorithms to generate candidates, while the ranking stage applies learned models to select the optimal route based on user preferences and context. Within this paradigm, R4 \cite{cheng2021r4} learns route representations via edge-level ID embeddings but encounters scalability issues due to the vast number of road segments in large networks. DSFNet \cite{yu2025dsfnet} introduces scenario-specific parameter decomposition for multi-scenario route ranking. However, existing route recommendation methods operate on item-level route representations and do not model the comparison-level information between route pairs that is central to distinguishing similar candidates.

\subsection{Pairwise Interaction in Recommendation}
Modeling interactions among candidate items is fundamental to recommendation quality. Attention-based methods \cite{zhou2018deep, zhou2019deep, kang2018self, sun2019bert4rec, chen2019behavior, chang2023twin} capture user preference patterns through target attention or self-attention over behavioral sequences. In listwise re-ranking, methods such as PRM \cite{pei2019personalized}, DLCM \cite{ai2018learning}, and Seq2Slate \cite{bello2018seq2slate} encode mutual influence among all candidates to optimize the overall ranking \cite{ren2024non}. Generative recommendation approaches \cite{rajput2023recommender, deng2025onerec} further model item sequences autoregressively to predict next interactions. Despite their diversity, these methods share a common item-first paradigm: they compute individual item representations and derive pairwise relations indirectly through attention weights or score differences. Classic pairwise learning methods such as BPR \cite{rendle2009bpr} explicitly model pairwise preferences but still operate on independently computed item scores without cross-pair interaction. In contrast, CCN constructs explicit comparison features from non-overlapping route segments as primary input and reasons directly in the pairwise space, where each pair's representation is informed by comparisons against all other candidates.

\section{Method}
\subsection{Overview}
\label{section:model_overview}
Given a candidate route set $\mathcal{R} = \{r_1, \ldots, r_N\}$ produced by recall and coarse ranking, where $N$ is typically small, e.g., $\sim$125 online and $\leq$30 in training, the ranking stage learns a scoring function over $\mathcal{R}$ conditioned on route attributes, user preferences, and contextual information to select the optimal route.

\begin{figure}[t]
\centering
    \includegraphics[width=0.48\textwidth]{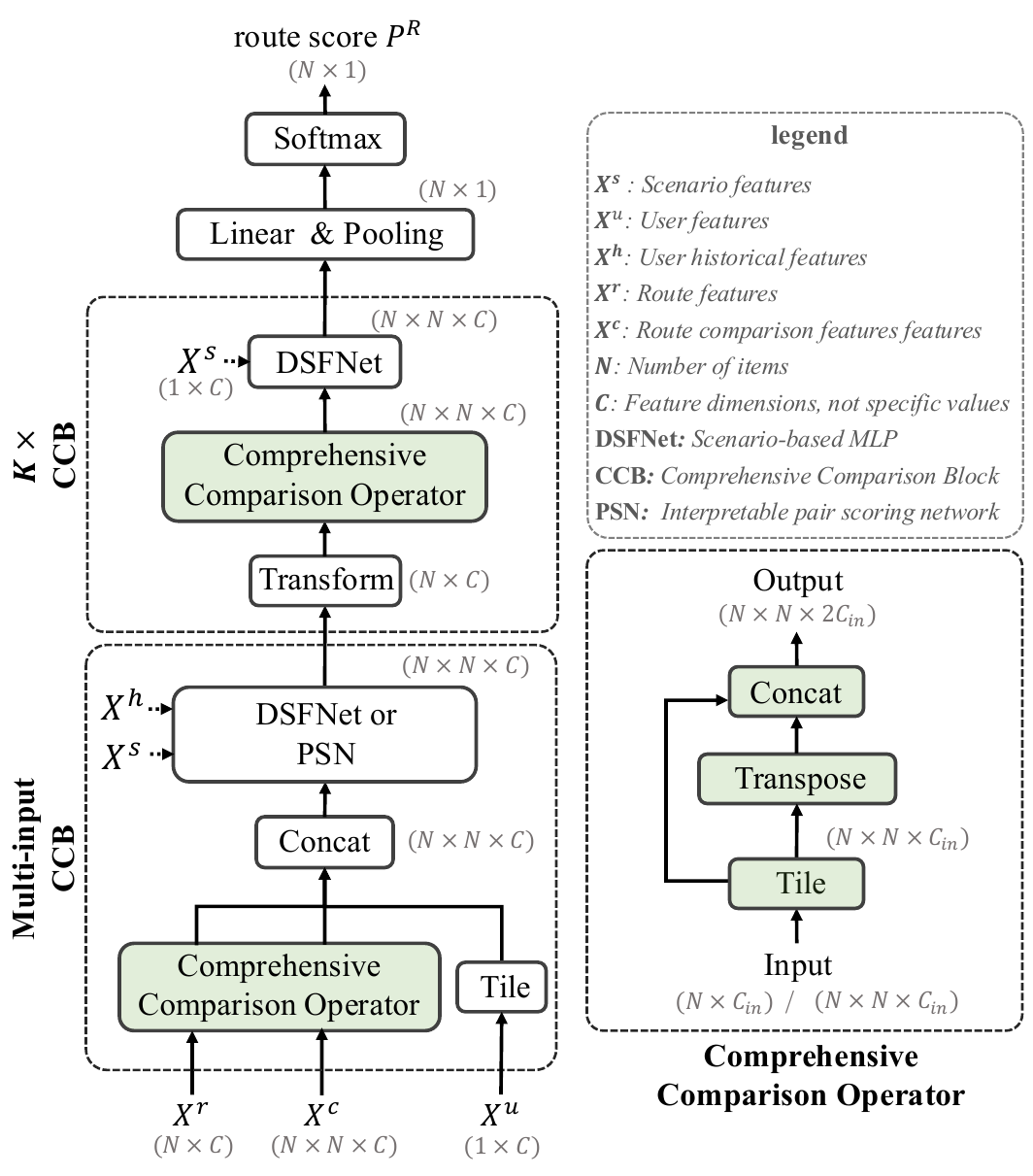}
\caption{Architecture of CCN. Left: $K$ stacked CCBs progressively refine route representations through CCO and DSFNet; the first CCB accepts multi-level inputs and optionally uses PSN for interpretability. Right: the CCO operator.}
\Description{Architecture diagram of CCN, including stacked Comprehensive Comparison Blocks, the Comprehensive Comparison Operator, multi-level route and user inputs, and optional Pair Scoring Network for interpretable pairwise scoring.}
\label{fig:2}
\end{figure}

As shown in Figure~\ref{fig:2}, CCN is a listwise ranking model built from $K$ stacked Comprehensive Comparison Blocks (CCBs). Each CCB applies a Comprehensive Comparison Operator (CCO) to organize features into pairwise representations, followed by a DSFNet~\cite{yu2025dsfnet} for multi-scenario scoring. The first CCB takes route features $X^r$, user profile and historical preference features, and \mbox{comparison-level} features $X^c$ constructed from non-overlapping route segments (Section~\ref{sec:comparison_feature}), and its DSFNet can be replaced by a Pair Scoring Network (PSN) for field-level interpretability (Section~\ref{sec:psn}). The final route score is obtained via linear projection and softmax.

\subsection{Comparison Feature Construction}
\label{sec:comparison_feature}
As illustrated in Figure~\ref{fig:1}(b), candidate routes between the same origin and destination typically share some road segments and diverge on others. We define the non-overlapping segments of route $i$ with respect to route $j$ as $\text{Route}_{i\text{-}j}$, i.e., the segments present in route $i$ but absent in route $j$. The \mbox{comparison-level} feature $X^c_{ij} \in \mathbb{R}^{F}$ is computed by aggregating physical attributes (e.g., distance, ETA, road grade) over $\text{Route}_{i\text{-}j}$. Repeating this for all ordered pairs yields the comparison tensor $X^c \in \mathbb{R}^{N \times N \times F}$.

$X^c_{ij}$ captures information that the item-level feature difference $X^r_i - X^r_j$ cannot. In Figure~\ref{fig:1}(b), item-level features only reveal that routes 2 and 3 are both 1\,km longer than route 1, making them indistinguishable. \mbox{Comparison-level} features, however, capture \emph{where} the extra distance occurs: route 2 diverges onto a spatially diverse path, whereas route 3 detours locally. Feature differencing loses this distinction because it aggregates over the entire route and discards spatial location.

This construction operates at the granularity of non-overlapping segments, preserving spatial difference information lost by full-route aggregation while avoiding the variable-length alignment and sparsity challenges of segment-level sequences. Two properties follow directly. \textbf{Asymmetry}: $X^c_{ij} \neq X^c_{ji}$, since the non-overlapping segments of route $i$ w.r.t.\ $j$ differ from those of $j$ w.r.t.\ $i$. \textbf{Pair-dependence}: $X^c_{ij} \neq X^c_{ik}$ for $j \neq k$, since the reference route determines which segments are non-overlapping. Together, $X^c$ encodes $N(N{-}1)$ distinct directed pairwise views. We acknowledge that this aggregation has a lower theoretical information capacity than segment-level sequence representations or learned ID embeddings. However, those alternatives demand more complex architectures, larger training data, and longer convergence time with no guaranteed improvement, making the proposed construction a practical optimum balancing expressiveness and deployability.

\begin{figure}[t]
\centering
\includegraphics[width=0.48\textwidth]{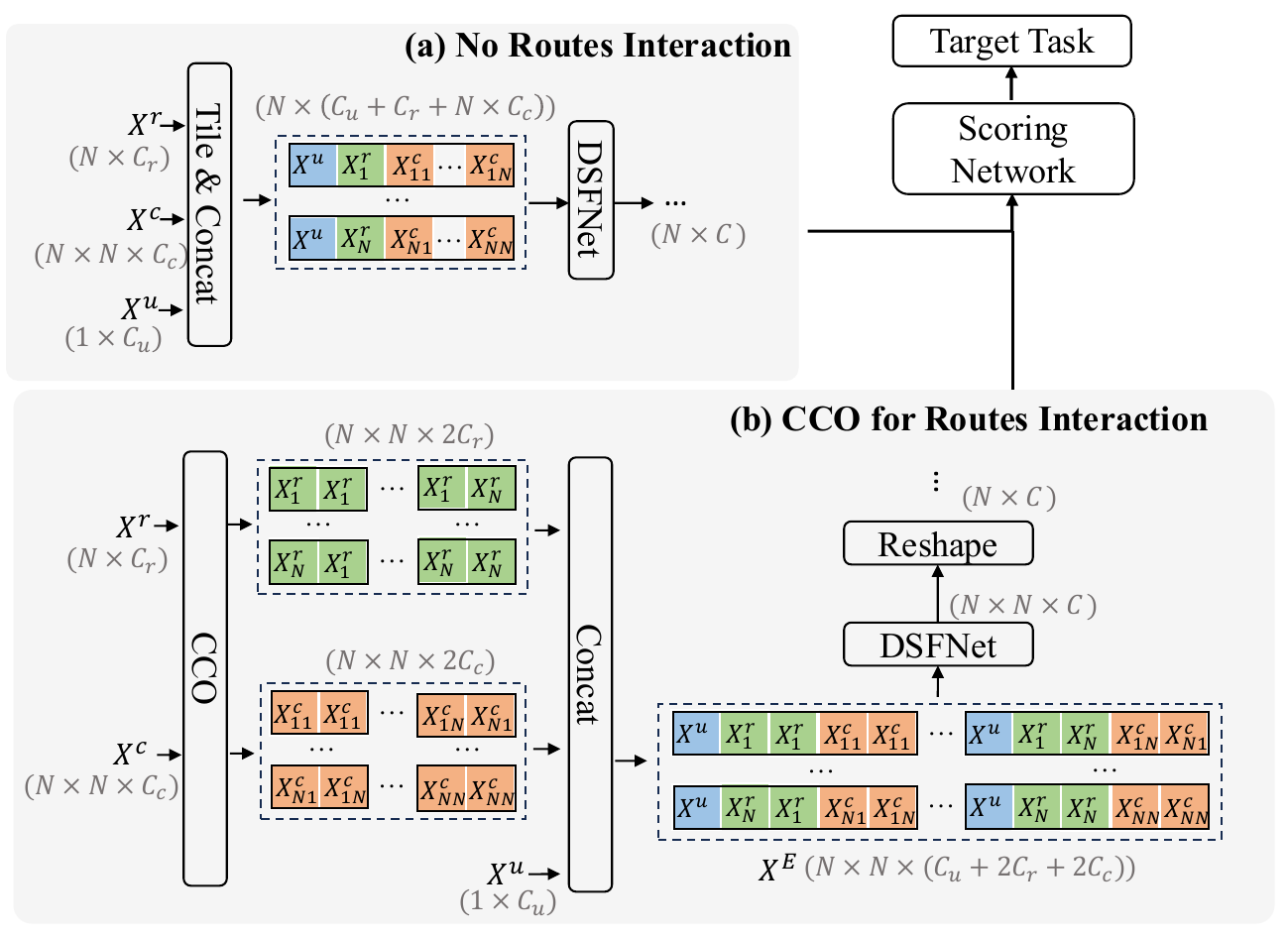}
\caption{(a) Without CCO, features are tiled and concatenated into item-level representations. (b) CCO organizes $X^r$ and $X^c$ into pairwise form, concatenates with $X^u$ to form $X^E$, then scores via DSFNet.}
\Description{Comparison of feature organization with and without CCO, showing how route features and comparison-level features are converted into pairwise representations and combined with user features for scoring.}
\label{fig:cco}
\end{figure}

\subsection{Comprehensive Comparison Block}
\label{sec:ccb}
CCB is the core module of CCN that organizes features into pairwise representations and enables cross-candidate reasoning through iterative stacking. Each CCB consists of a Comprehensive Comparison Operator (CCO) for pairwise feature organization and an interaction network for capturing competitive context among candidates.

\textbf{Comprehensive Comparison Operator (CCO).}
Evaluating whether route $i$ is preferable to route $j$ requires simultaneous access to the features of both routes. CCO achieves this by organizing input features into pairwise representations (Figure~\ref{fig:cco}). For item-level input $X \in \mathbb{R}^{N \times C}$:
\begin{equation}  
\text{CCO}(X)_{ij} = [X_i \| X_j] \in \mathbb{R}^{2C},
\label{eq:cco_item}
\end{equation}
and for pairwise input $X \in \mathbb{R}^{N \times N \times C}$:
\begin{equation}  
\text{CCO}(X)_{ij} = [X_{ij} \| X_{ji}] \in \mathbb{R}^{2C},
\label{eq:cco_pair}
\end{equation}
where $\|$ denotes concatenation. The first form applies to $X^r$ in the first CCB, and the second form applies to $X^c$ in the first CCB and to pairwise outputs in subsequent CCBs.

\textbf{First CCB: multi-source input assembly.}
The first CCB applies CCO to route features $X^r$ and comparison features $X^c$ separately, then concatenates the results with user features $X^u$ to form the pairwise representation:
\begin{equation}
X^E_{ij} = [\text{CCO}(X^r)_{ij} \| \text{CCO}(X^c)_{ij} \| X^u] \in \mathbb{R}^{C},
\label{eq:xe}
\end{equation}
where $X^u$ denotes user profile features. The assembled $X^E \in \mathbb{R}^{N \times N \times C}$ is then scored by DSFNet~\cite{yu2025dsfnet}:
\begin{equation}
P^E = \text{DSFNet}(X^E, X^s) \in \mathbb{R}^{N \times N \times C},
\label{eq:mccb_2}
\end{equation}
where $X^s$ denotes scene features. DSFNet is a multi-scenario scoring network that adapts its parameters to the current scenario through linear decomposition. Given scene feature $X^s \in \mathbb{R}^{1 \times F}$, it computes scenario-specific weights:
\begin{equation}
\alpha^{(l)}= \sigma(X^s W^{(l)}_s),
\label{eq:dsfnet1}
\end{equation}
where $W^{(l)}_s \in \mathbb{R}^{F \times D}$ maps the scene feature to $D$ scenario-specific coefficients at layer $l$, and $\sigma$ is the sigmoid function. The layer output is then computed as:
\begin{equation}
\hat{X}^{(l)} = \left(\sum_{i=1}^{D} \alpha_i \mathbf{W}^{(l)}_i\right) X^{(l)},
\label{eq:dsfnet2}
\end{equation}
where $X^{(l)}$ denotes the input of layer $l$ and $\mathbf{W}^{(l)}_i$ is the projection matrix for the $i$-th learner. In this work, all DSFNets are four-layer fully connected networks with batch normalization. The DSFNet can also be replaced by a Pair Scoring Network for field-level interpretability, as detailed in Section~\ref{sec:psn}.

\textbf{Stacked CCBs: propagating cross-candidate context.}
A single CCB captures only direct pairwise comparisons. To enable each pair's representation to be informed by how both routes compare against all other candidates, we stack $K$ CCBs. Each subsequent CCB first aggregates pairwise output back to item-level by mean-pooling over comparison partners, then re-expands via CCO and scores again:
\begin{equation}
X^k_i = \frac{1}{N}\sum_{j=1}^{N} O^{k-1}_{ij}, \quad O^k = \text{DSFNet}(\text{CCO}(X^k), X^s),
\label{eq:ccb_stack}
\end{equation}
for $1 < k \le K$, where $O^0 = P^E$ and $X^k \in \mathbb{R}^{N \times C}$ summarizes how each route compares against the full candidate set. After $K$ layers, the final route score is:
\begin{equation}
P^R = \text{Softmax}\left(\frac{1}{N}\sum_{j=1}^{N} (O^K W)_{ij}\right) \in \mathbb{R}^{N \times 1},
\label{eq:route_score}
\end{equation}
where $W \in \mathbb{R}^{C \times 1}$ is a learnable projection.

\begin{figure}[t]
\centering
    \includegraphics[width=0.48\textwidth]{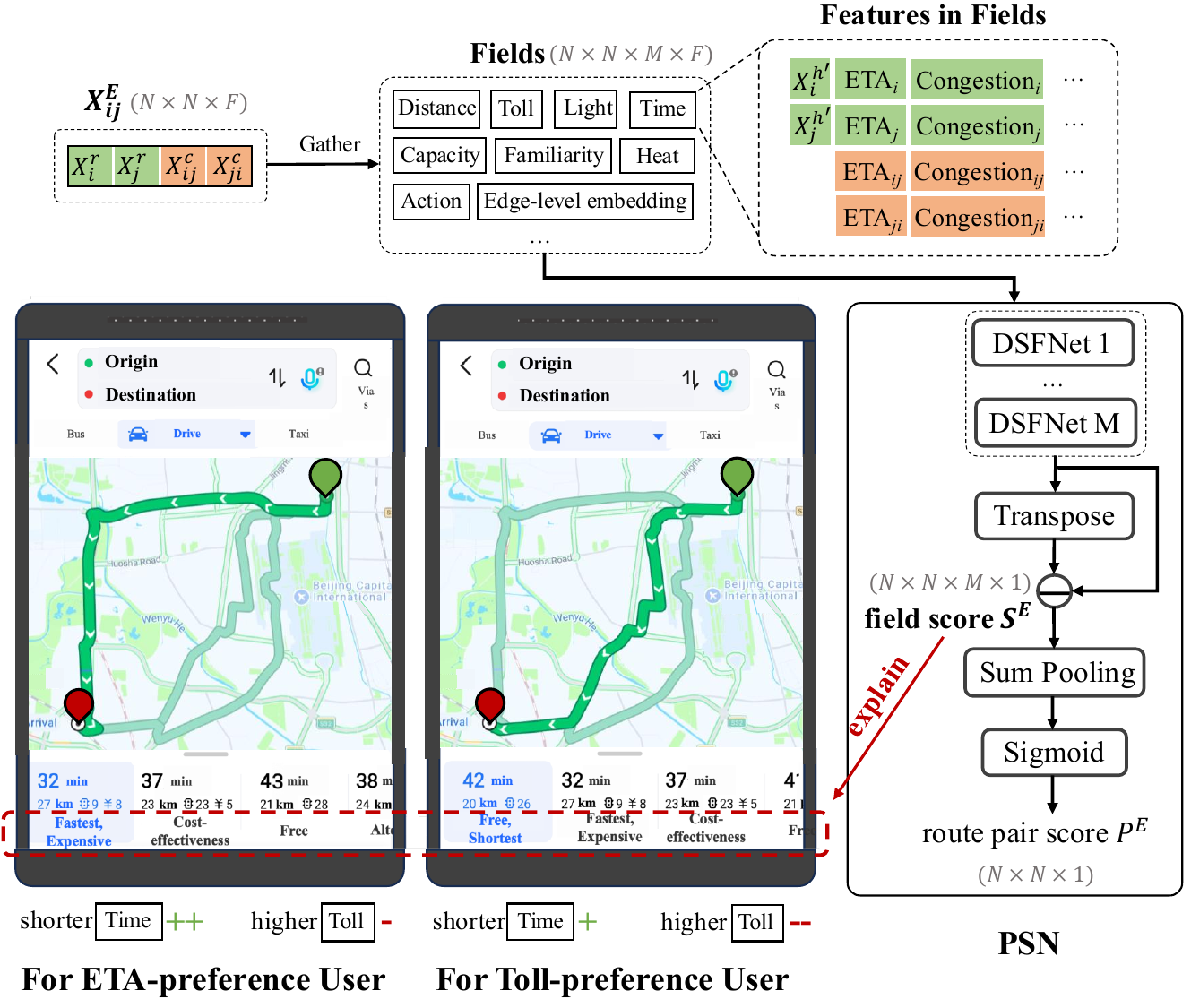}
\caption{Overview of PSN. Left: Gather partitions $X^E$ into $M$ semantic fields and inserts $X^h$ into each route position. Right: each field is scored by a dedicated DSFNet, producing interpretable field scores $S^E$. Bottom: an ETA-preference user and a toll-preference user receive different orderings from the same candidate set.}
\Description{Overview of the Pair Scoring Network, showing field-wise decomposition of pairwise features, field-specific scoring networks, and examples where different user preferences lead to different route orderings.}
\label{fig:field}
\end{figure}

\begin{table*}[!t]
  \centering
  \caption{Feature categories and representative attributes in our dataset.}
  \setlength{\tabcolsep}{8pt}
  \begin{tabular}{@{}l|c|l@{}}
    \toprule
    Feature Category & Dimension & Representative Attributes \\
    \midrule
    \multirow{4}{*}{Route Features} & \multirow{4}{*}{$N\times 62$} & Estimated time of arrival (ETA)\\
    ~&~& Total distance\\
    ~&~& Total toll cost\\
    ~&~& Number of traffic lights\\
    \midrule
    \multirow{2}{*}{Scene Features} & \multirow{2}{*}{$1\times 10$} & Destination POI category \\
    ~&~& Navigation request time \\
    \midrule
    User Profiles & $1\times 12$ & Driving behavior characteristics \\
    \midrule
    Historical Preferences & $1\times 77$ & Familiar routes and habitual route choices \\
    \midrule
    \multirow{2}{*}{Comparison-Level Features} & \multirow{2}{*}{$N\times N \times 27$} & ETA difference of non-overlapping segments \\
    ~&~& Distance difference of non-overlapping segments \\
    \bottomrule
  \end{tabular}
  \label{tab:datafeat}
\end{table*}

\subsection{Pair Scoring Network}
\label{sec:psn}
To provide field-level interpretability for online deployment, we design a Pair Scoring Network (PSN) that replaces DSFNet in the first CCB. PSN decomposes $X^E$ into $M$ fields according to physical attribute semantics (e.g., ETA, toll, road grade):
\begin{equation}
X^E_m = \text{Gather}(X^E, X^h, m), \quad \forall m,
\end{equation}
where Gather partitions $X^E$ by attribute semantics and inserts the full user historical preferences $X^h$ into each route position within every pair. $X^h$ encodes long-term route selection patterns and is broadcast without decomposition, allowing each field-specific network to learn relevant preference dimensions and capture cross-preference interactions. Each field is then scored by a dedicated network:
\begin{equation}
\hat{X}_m = \text{DSFNet}_{m}(X^E_m, X^s) \in \mathbb{R}^{N \times N \times 1}, \quad \forall m.
\label{eq:psn_1}
\end{equation}
To enforce antisymmetry in pairwise preferences, PSN computes relative field scores:
\begin{equation}
S^E_{ij,m} = \hat{X}_{ij,m} - \hat{X}_{ji,m},
\label{eq:psn_3}
\end{equation}
where $S^E_{ij,m}$ indicates the preference of route $i$ over route $j$ in field $m$, providing transparent per-field explanations as illustrated in Figure~\ref{fig:field}. The final pairwise score aggregates across fields:
\begin{equation}
P^E_{ij} = \sigma\left(\sum_{m=1}^{M} S^E_{ij,m}\right),
\label{eq:psn_4}
\end{equation}
where $\sigma$ is the sigmoid function. Unlike post-hoc attribution methods~\cite{lundberg2017unified,sundararajan2017axiomatic} that approximate feature importance after training, PSN produces field scores as explicit intermediate outputs during the forward pass. The additive aggregation across fields ensures that each $S^E_{ij,m}$ is a causally faithful contribution to the final decision. This design serves two purposes: user-facing route labels that explain why a route is recommended, and engineering-facing diagnostics that enable query-level bad case analysis and feature quality monitoring in production.

\begin{table*}[!t]
\centering
\caption{Comparison results of different methods on the MSDR test set. RelaImpr is calculated based on the vanilla MLP, where RelaImpr represents the relative improvement of AUC (\%). The best results are shown in boldface and the second best are underlined.}
\scalebox{1.05}{
\begin{tabular}{l|cc|cc|cc|cc|cc}
\toprule
Method & AUC$_{1}^{11}$ & RelaImpr & AUC$_{12}^{36}$ & RelaImpr & AUC$_{37}^{61}$ & RelaImpr & AUC$_{62}^{72}$ & RelaImpr & AUC & RelaImpr \\
\midrule
Recall & 50.36 & -31.75\% & 48.99 & -33.77\% & 52.17 & -29.41\% & 55.77 & -22.80\% & 50.16 & -32.08\%  \\
Vanilla MLP & 73.55 & 0.00\% & 73.71 & 0.00\% & 73.70 & 0.00\% & 72.09 & 0.00\% & 73.62 & 0.00\% \\
ShareBottom~\cite{caruana1997multitask} & 74.17 & 0.85\% & 73.79 & 0.11\% & 73.66 & -0.05\% & 68.10 & -5.57\% & 74.08 & 0.63\% \\
MMOE~\cite{ma2018modeling} & 74.18 & 0.86\% & 73.96 & 0.34\% & 73.65 & -0.07\% & 68.17 & -5.48\% & 74.15 & 0.72\% \\
PLE~\cite{tang2020progressive}  & 74.21 & 0.90\% & 73.93 & 0.30\% & 73.87 & 0.23\% & 68.43 & -5.11\% & 74.16 & 0.74\% \\
STAR~\cite{sheng2021one}  & 74.18 & 0.86\% & 74.01 & 0.41\% & 73.72 & 0.03\% & 69.54 & -3.56\% & 74.13 & 0.70\% \\
M2M~\cite{zhang2022leaving} & 73.57 & 0.03\% & 74.04 & 0.45\% & 74.17 & 0.64\%  & 70.64 & -2.03\% & 73.67 & 0.07\% \\
APG~\cite{yan2022apg} & 72.89 & -0.90\% & 72.58 & -1.54\% & 73.37 & -0.45\% & 70.92 & -1.63\% & 72.91 & -0.97\% \\
HiNet~\cite{zhou2023hinet} & 74.35 & 1.10\%  & 74.65 & 1.28\% & 74.46 & 1.04\% & 70.41 & -2.35\% & 74.41 & 1.08\% \\
MuSeNet~\cite{xu2023musenet} & 74.56 & 1.23\% & 74.23 & 0.71\% & 75.08 & 1.89\% & 71.40 & -0.96\% & 74.52 & 1.23\% \\
DSFNet~\cite{yu2025dsfnet} & 74.96 & 1.93\%  & 75.72 & 2.75\%  & \underline{75.86} & \underline{2.95\%} & 72.20 & 0.15\% & 75.04 & 1.94\% \\
\rowcolor{gray!15}
DSFNet-SA & \underline{75.27} & \underline{2.34\%} & \underline{75.91} & \underline{2.98\%} & 75.76 & 2.80\% & \underline{75.39} & \underline{4.58\%} & \underline{75.87} & \underline{3.06\%} \\
\rowcolor{gray!15}
CCN (w/o PSN, $X^c$) & \textbf{76.67} & \textbf{4.24\%} & \textbf{76.32} & \textbf{3.54\%} & \textbf{76.40} & \textbf{3.66\%} & \textbf{75.40} & \textbf{4.59\%} & \textbf{76.72} & \textbf{4.21\%} \\
\bottomrule
\end{tabular}
}
\label{table:comparison_results_MSDR}
\end{table*}

\subsection{Training Objective}
\label{sec:training}
CCN outputs a single recommended route from the candidate set. The primary objective is an $N$-way classification loss over the route selection probability $P^R$, weighted by the label quality:
\begin{equation}
L_{R}=-w_l \cdot Y^R \log P^R,
\label{eq:loss_1}
\end{equation}
where $Y^R$ is a one-hot label assigned to route $l$, the candidate with the highest overlap against the user's actual travel trajectory. The scalar $w_l \in [0,1]$ is this overlap ratio. A high $w_l$ indicates that the label reliably reflects user intent and deserves full gradient contribution. A low $w_l$ means no candidate closely matches the actual trajectory, so the training signal is down-weighted to suppress label noise.

To provide explicit pairwise supervision for the field scores of PSN, we introduce an auxiliary loss $L_P$. Using route $l$ as the anchor, we assign $Y^E_{lj}=1$ and $Y^E_{jl}=0$ for all $j \neq l$, meaning the highest-overlap route is preferred over every other candidate. Only pairs involving $l$ are supervised, as other pairs lack reliable ground truth. The pairwise loss is:
\begin{equation}
L_{P}=w_l \cdot \sum_{i=1}^N\sum_{j=1}^N \mathbb{I}_{(i=l\lor j=l) \land i\neq j} \cdot (-Y^E_{ij}\log P^E_{ij}),
\label{eq:loss_p}
\end{equation}
where $P^E_{ij} \in [0,1]$ is the predicted pair score from PSN. Through this supervision, PSN learns to produce field scores consistent with actual user preferences. Specifically, $S^E_{ij,m} > 0$ indicates that route $i$ is preferred over route $j$ in field $m$. This interpretable semantics is directly guaranteed by the training objective rather than post-hoc attribution.

The total loss combines both terms:
\begin{equation}
L=L_R + \lambda\cdot L_P,
\label{eq:loss_total}
\end{equation}
where $\lambda$ controls the contribution of the pairwise auxiliary loss.

\section{The Proposed Dataset}
\label{sec:dataset}

\subsection{Dataset Construction}
We target the pre-navigation route recommendation task. Given a destination specified by the user, the system recalls $N$ candidate routes via planning algorithms~\cite{hart1968formal, abraham2013alternative}, and a ranking model selects the optimal route to present. To support this task, we collect data from AMAP, serving over 175 million users across 370 cities. The dataset comprises 512 million samples and 6 billion routes. The training set uses $N < 30$ routes per sample, filtered from over 100 candidates to retain diverse route options. Each sample includes 62-dimensional route features, scene features, user profiles, user historical preferences, and comparison-level features proposed in this work. The ground-truth label is constructed from the actual user trajectory. We compute the overlap ratio between the trajectory and each recalled route, and assign the route with the highest overlap as the positive label. This trajectory-based labeling provides more reliable supervision than click-based approaches~\cite{yu2025dsfnet}, as it reflects actual driving behavior. Table~\ref{tab:datafeat} presents examples of key features in the dataset.

\subsection{Evaluation Metrics}
\textbf{Coverage Rate (CR).} CR measures the overlap between a recommended route $R$ and the actual user trajectory $T$:
\begin{equation}
\text{CR} = \frac{|R \cap T|}{|R \cup T|}.
\label{eq:CR}
\end{equation}
Higher CR indicates better alignment with user intent. We use CR for both offline and online evaluation.

\textbf{Deviation Rate (DR).} DR measures the probability of users deviating from the recommended route during navigation:
\begin{equation}
\text{DR} = \frac{D}{R_{\text{total}}},
\end{equation}
where $D$ denotes the number of deviations and $R_{\text{total}}$ denotes the total number of recommendations. Lower DR indicates better performance. DR is available only in online experiments as deviation labels require real-time navigation data.

\textbf{Pair Score Accuracy (ACC).} To evaluate the quality of PSN pair scores, we compute ACC between the highest-CR route $l$ and all other routes:
\begin{equation}
\text{ACC} = \frac{1}{|S|} \sum_{(i,j) \in S} \mathbb{I}\left[(P^{E}_{ij} > 0.5) = Y^{E}_{ij}\right],
\label{eq:psn_acc}
\end{equation}
where $S = \{(i,j) \mid (i=l \lor j=l) \land i \neq j\}$. Higher ACC indicates that PSN field scores faithfully reflect actual user preferences.

\section{Experiments}
\subsection{Experimental Settings}
\label{sec:Exper_details}
We evaluate CCN on two datasets. The first is our proposed dataset described in Section~\ref{sec:dataset}. The second is MSDR~\cite{yu2025dsfnet}, a public route recommendation dataset containing 59 million samples across 8 cities, designed for predicting user clicks on replanned routes during navigation deviations. For MSDR, following DSFNet, we apply AUC and RelaImpr as metrics, with subset AUC scores $\text{AUC}^{j}_{i}$ computed for scenario ranges partitioned by sample frequency.

CCN is trained on 8 NVIDIA H20 GPUs with a batch size of 128 and a learning rate of 0.001 using the Adam optimizer. The number $K$ of stacked CCB layers is set to 3 and the loss weighting parameter $\lambda$ is set to 1.0. The model converges at approximately 1.2 million global steps on our proposed dataset. For MSDR, we use the same configuration except for the number of training steps, adjusted according to dataset size.

\subsection{Comparison with Various Methods}
\subsubsection{Results on MSDR}
Table~\ref{table:comparison_results_MSDR} presents the comparison results on the MSDR test set. \textbf{Recall} denotes routes with the highest expert scores during recall. \textbf{Vanilla MLP} is a naive multi-layer perceptron. \textbf{DSFNet-SA} augments DSFNet with self-attention for route interaction. Since MSDR lacks paired route comparison data required to construct comparison-level features $X^c$, we evaluate \textbf{CCN} (w/o PSN, $X^c$) to isolate the contribution of the Comprehensive Comparison Block alone. Even without comparison-level features and PSN, CCN achieves consistent AUC improvements over both DSFNet and the transformer-based DSFNet-SA across all scenario partitions, with an overall AUC gain of 1.68\% over DSFNet. This confirms that context-aware pairwise reasoning in CCB, where each pair comparison is informed by the full candidate set, provides effective route modeling even when only item-level features are available.

\subsubsection{Results on the Proposed Dataset}
Table~\ref{table:comparison_results_our} presents the results on our proposed dataset. \textbf{SD} and \textbf{ST} denote the shortest distance and shortest time algorithms. \textbf{Recall} denotes routes with the highest expert scores during recall. \textbf{DSFNet-SA} augments DSFNet with self-attention and comparison-level features as shown in Figure~\ref{fig:cco}(b). \textbf{CCN} is our full model with comparison-level features, CCB, and PSN. CCN achieves the best performance across all metrics, including 85.70\% offline CR and +1.2\% online CR improvement with -0.8\% deviation rate reduction. The improvement over DSFNet-SA demonstrates that explicitly modeling pairwise comparisons through structured comparison features and context-aware reasoning captures route differences more effectively than attention-based item-level interaction. The comparison-level features preserve segment-level spatial information lost by item-level aggregation, and CCB further leverages the full competitive context among all candidates to refine pairwise judgments.

\begin{table}[t]
\centering
\caption{Comparison with various methods on the proposed dataset. Bold and underline denote the best and second best results. $\rm CR_{on}$ and DR are relative to DSFNet.}
\scalebox{1.0}{
\begin{tabular}{l@{\hskip 12pt}c@{\hskip 12pt}c@{\hskip 12pt}c}
\toprule
Model & ${\rm CR_{off}}$ (\%) & ${\rm CR_{on}}$ (\%) & ${\rm DR}$ (\%) \\
\midrule
SD & 64.34 & -20.49 & +10.26  \\
ST  & 73.80 & -10.72 & +4.72  \\
Recall & 75.63 & -9.11 & +4.20   \\
DSFNet  & 83.70 & +0 & +0   \\ 
\rowcolor{gray!15}
DSFNet-SA  & \underline{84.27} & \underline{+0.8} & \underline{-0.02}   \\
\rowcolor{gray!15}
CCN (Ours) & \textbf{85.70} & \textbf{+1.2} & \textbf{-0.8}  \\
\bottomrule
\end{tabular}
}
\label{table:comparison_results_our}
\end{table}

\subsection{Ablation Study}
\subsubsection{Ablation Analysis of CCN Components}
Table~\ref{table:ablation_1} presents the ablation results. We analyze the contribution of each component from three perspectives.

\textbf{Comparison-level features and CCO drive recommendation performance.} The configuration with $X^c$ and CCO but without PSN in row 5 improves offline CR from 83.70\% to 85.67\% and online CR by +1.08\% over the baseline in row 1. This confirms that explicit comparison input combined with context-aware pairwise reasoning accounts for the majority of performance gain, independent of PSN.

\textbf{$X^c$ and CCO are essential for PSN interpretability quality.} PSN alone in row 2 achieves only 91.71\% ACC and degrades online CR by -4.96\%, as field-level scoring without sufficient pairwise signals lacks discriminative power. Adding $X^c$ in row 3 raises ACC to 93.13\% by providing explicit comparison signals for each field. Adding CCO in row 4 further raises ACC to 93.39\% by incorporating the full candidate set context into pairwise judgments. Both components supply the structured information that PSN requires to produce meaningful field scores.

\textbf{PSN provides additional online gains when supported by $X^c$ and CCO.} Comparing the full model in row 6 against the configuration without PSN in row 5, offline CR remains comparable at 85.70\% versus 85.67\%, but the deviation rate improves from -0.4\% to -0.8\%. This indicates that field-level reasoning in PSN, when grounded in high-quality comparison signals, further aligns recommendations with actual user navigation behavior, yielding measurable online improvements.

\begin{table}[t]
\centering
\caption{Ablation analysis of CCN components. $X^c$, CCO, and PSN denote comparison-level features, comprehensive comparison operator, and pair scoring network, respectively.}
\setlength{\tabcolsep}{4pt}
\begin{tabular}{ccc|cccc}
\toprule
$X^c$ & CCO & PSN & ${\rm CR_{off}}$ (\%) & ACC (\%) & ${\rm CR_{on}}$ (\%) & ${\rm DR}$ (\%)\\
\midrule
$\times$ & $\times$ & $\times$ & 83.70 & - & +0 & +0\\
$\times$ & $\times$ & $\checkmark$ & 83.20 & 91.71 & -4.96 & +3.64 \\
$\checkmark$ & $\times$ & $\checkmark$ & 84.94 & 93.13 & +0.36 & +0.12 \\
$\times$ & $\checkmark$ & $\checkmark$ & 84.71 & \underline{93.39} & +0.08 & +0.2\\
$\checkmark$ & $\checkmark$ & $\times$ & \underline{85.67} & - & \underline{+1.08} & \underline{-0.4}\\
\rowcolor{gray!15}
$\checkmark$ & $\checkmark$ & $\checkmark$ & \textbf{85.70} & \textbf{93.66} & \textbf{+1.2} & \textbf{-0.8} \\
\bottomrule
\end{tabular}
\label{table:ablation_1}
\end{table}

\begin{table}[t]
\centering
\caption{Impact of the number $K$ of stacked CCB layers. CR denotes offline coverage rate and ACC denotes PSN pair score accuracy.}
\setlength{\tabcolsep}{6pt}
\begin{tabular}{l|cccccc}
\toprule
$K$ & 1 & 2 & \textbf{3} & 4 & 5 & 6 \\
\midrule
CR (\%) & 85.43 & 85.68 & \textbf{85.70} & 85.68 & 85.66 & 85.65 \\
ACC (\%) & 93.58 & 93.65 & 93.66 & \textbf{93.67} & \textbf{93.67} & \textbf{93.67} \\
\bottomrule
\end{tabular}
\label{table:appendix_ablation_2}
\end{table}

\begin{table*}[t]
\centering
\caption{Sensitivity analysis of the loss weighting parameter $\lambda$. CR denotes offline coverage rate and ACC denotes PSN pair score accuracy. Performance remains stable across the full range.}
\setlength{\tabcolsep}{6pt}
\begin{tabular}{l|ccccccccccc}
\toprule
$\lambda$ & 0.0 & 0.1 & 0.2 & 0.3 & 0.4 & 0.5 & 0.6 & 0.7 & 0.8 & 0.9 & \textbf{1.0} \\
\midrule
CR (\%) & 85.67 & 85.69 & 85.70 & 85.71 & 85.73 & 85.74 & \textbf{85.75} & \textbf{85.75} & 85.71 & 85.70 & 85.70 \\
ACC (\%) & - & 93.64 & 93.65 & 93.65 & 93.65 & 93.66 & \textbf{93.67} & 93.66 & 93.66 & 93.66 & 93.66 \\
\bottomrule
\end{tabular}
\label{table:appendix_ablation_3}
\end{table*}

\begin{figure}[t]
\centering
    \includegraphics[width=0.495\textwidth]{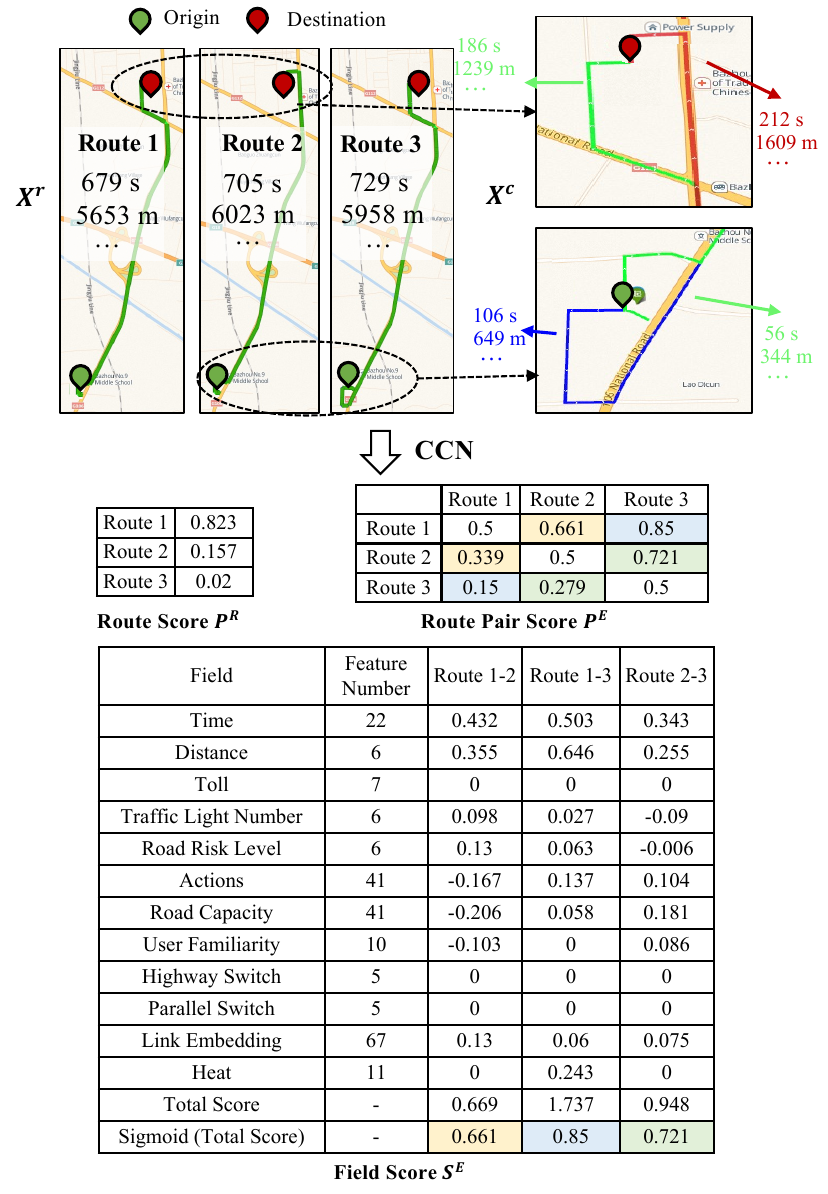}
\caption{A case study of PSN interpretability with three candidate routes. Top: route features $X^r$ and comparison-level features $X^c$. Middle: route scores $P^R$ and pairwise preference matrix $P^E$. Bottom: field-level scores $S^E$ decomposed by physical attributes, where positive values indicate the row route is preferred over the column route in that field.}
\Description{Case study visualizing three candidate routes, their route scores, pairwise preference matrix, and field-level preference scores that explain why the selected route is preferred.}
\label{fig:exp}
\end{figure}

\subsubsection{Impact of the Number $K$ of Stacked CCB Layers}
Table~\ref{table:appendix_ablation_2} presents the impact of varying $K$. Increasing $K$ progressively enriches the competitive context available to each pairwise representation. Performance peaks at $K=3$ with 85.70\% offline CR. Beyond this point, CR slightly declines due to overfitting on the fixed-size candidate set. ACC remains stable across all $K$ values, as it is directly optimized by the pairwise loss and largely independent of stacking depth. We also note that deeper stacking may slightly reduce the fidelity of PSN field scores, since the final route selection incorporates higher-order interactions beyond the first-layer pairwise scoring. In practice, this effect is minor and does not materially affect interpretability quality. Balancing recommendation performance, computational cost, and interpretability, we select $K=3$.
\subsubsection{Impact of the Loss Weighting Parameter $\lambda$}
Table~\ref{table:appendix_ablation_3} presents the sensitivity analysis of $\lambda$. Both CR and ACC remain stable across the entire range, with CR varying within 0.08\% and ACC within 0.03\%. This demonstrates that CCN is insensitive to the choice of $\lambda$, eliminating the need for hyperparameter search in practice. When $\lambda=0$, PSN receives no gradient and ACC is undefined, yet CR remains at 85.67\%, confirming that PSN does not compromise recommendation performance. We select $\lambda=1.0$ as the default value, which ensures sufficient gradient signal for PSN convergence while maintaining competitive recommendation performance.

\subsection{Interpretability Analysis}
Figure~\ref{fig:exp} presents a case study with three candidate routes. We analyze PSN outputs at three levels of granularity.

\textbf{Route score $P^R$.} The final scores of 0.823, 0.157, and 0.02 identify Route~1 as the recommended route but provide no explanation for why it is preferred.

\textbf{Pairwise score $P^E$.} The preference of Route~1 over Route~3 reaches $P^E_{13}=0.85$, notably stronger than $P^E_{12}=0.661$ over Route~2. This reveals that Route~3 is a substantially weaker alternative, but the source of this difference remains opaque at the pairwise level.

\textbf{Field score $S^E$.} Decomposing the pairwise preference into physical fields reveals interpretable trade-offs. For the Route~1 vs.\ Route~2 pair, the Time field scores 0.432 and Distance scores 0.355, indicating that Route~1 is preferred in both travel time and distance dimensions. In contrast, the Actions field scores $-$0.167 and Road Capacity scores $-$0.206, indicating that Route~2 is preferred in driving maneuver complexity and road width dimensions. For the Route~1 vs.\ Route~3 pair, the Heat field scores 0.243 and Distance scores 0.646, indicating that Route~1 holds substantial advantages in both traffic heat and distance dimensions. In production, field scores are mapped to user-facing route tags that explain why a route is recommended, and also serve as engineering diagnostics for feature quality monitoring and query-level bad case analysis. This three-level decomposition from $P^R$ through $P^E$ to $S^E$ demonstrates that PSN provides progressively finer explanations without sacrificing recommendation accuracy, as validated by the 93.66\% ACC in the ablation study.

\subsection{Computational Efficiency}
\label{sec:efficiency}
The pairwise representation in CCN introduces $O(N^2)$ complexity, which is on par with attention-based alternatives such as DSFNet-SA that also require $O(N^2)$ interaction computation. We analyze its practical feasibility under production constraints.

The computational complexity of CCN is $O(K \cdot N^2 \cdot (F \cdot h_1+h_1 \cdot h_2+h_2 \cdot h_3+h_3 \cdot h_4))$, where $K$ is the number of stacked CCBs, $N$ is the candidate set size, $F$ is the pairwise feature dimension, and $h_i$ denotes the number of neurons in each DSFNet layer. With $K=3$, $N \approx 100$ for online deployment, and $F \approx 200$, the total model size is approximately 2\,MB. Compared to DSFNet with $O(N)$ complexity and 1\,MB model size, CCN requires approximately three times the computation. Despite this overhead, CCN remains fully deployable: a single server with 16 CPUs and 10\,GB memory handles 200--300 QPS with an average response time of 15\,ms, well within the latency budget of real-time navigation services. For scenarios with larger candidate sets beyond 100 routes, a lightweight pre-ranking stage can filter candidates before CCN scoring. This does not degrade recommendation quality because the recall set typically contains many near-duplicate or clearly inferior routes that can be eliminated with simple heuristics, while the competitive candidates that require fine-grained comparison remain in the filtered set for CCN to evaluate.

\section{Conclusion}
We propose the Comprehensive Comparison Network, a comparison-first ranking framework that treats comparative judgment as the decision-making essence of route recommendation. CCN constructs an $N \times N \times F$ comparison tensor from non-overlapping segments between route pairs, preserving spatial differences that are irreversibly lost by item-level aggregation. The Comprehensive Comparison Block enables context-aware pairwise reasoning in which each comparison is informed by the full candidate set. The Pair Scoring Network decomposes pairwise preferences into independent physical fields, producing field-level scores that serve as both user-facing route tags and engineering diagnostics in production. We further contribute a large-scale route recommendation dataset to facilitate future research. Extensive experiments on both a public benchmark and our proposed dataset demonstrate that CCN achieves state-of-the-art recommendation performance with faithful interpretability. CCN has been deployed as the production ranking model in Amap for over two years, serving hundreds of millions of users daily.

\begin{acks}
The authors used Qwen 3.6 solely for grammar correction, sentence clarity, and academic writing polishing; it was not used for data processing, experiments, model design, analysis, or conclusions.
\end{acks}

\bibliographystyle{ACM-Reference-Format}
\bibliography{sample-base}


\end{document}